\documentclass[a4paper,10pt]{article}
 
\usepackage{graphicx}
\usepackage{caption}
\usepackage{subcaption}
\usepackage{wrapfig}
\usepackage{endnotes}
\usepackage{authblk}                                           
\usepackage{amssymb}
\usepackage{bbm}                                               
\usepackage{tensor}                                            
\usepackage{amsmath}
\usepackage[usenames,dvipsnames]{color}
\usepackage{setspace}                           
\usepackage{hyperref}
\usepackage{epigraph}
\usepackage[utf8x]{inputenc}
\usepackage[left= 1in,right=1in,top=1in,bottom=1in]{geometry}                                  

\hypersetup{
    colorlinks=true,         
    linkcolor=blue,          
    citecolor=red,        
    urlcolor=Violet             
}

\let\oldmarginpar\marginpar
\renewcommand\marginpar[1]{\oldmarginpar{\color{red}\raggedright\scriptsize #1}}

\def\lf {\ensuremath{\left}}
\def\rt {\ensuremath{\right}}

\title{{\bf Is Spacetime Countable?}}

\author{Written by {\bf Sean Gryb} and illustrated by {\bf Marc Ngui}}

\date{}                         

\begin{document}

\maketitle

\begin{abstract}
    Is there a number for every bit of spacetime, or is spacetime smooth like the real line? The ultimate fate of a quantum theory of gravity might depend on it. The troublesome infinities of quantum gravity can be cured by assuming that spacetime comes in countable, discrete pieces which one could simulate on a computer. But, perhaps there is another way? In this essay, we propose a picture where scale is meaningless so that there can be no minimum length and, hence, no fundamental discreteness. In this picture, Einstein's Special Relativity, suitably modified to accommodate an expanding Universe, can be reinterpreted as a theory where only the instantaneous shapes of configurations count.	
\end{abstract}
\clearpage 
 
\noindent ``Not everything that counts can be counted and not everything that can be counted counts.''
\begin{flushright} -- Albert Einstein \end{flushright}

\section{Counting What Counts}

This essay is about what things we can count, and what we can't. Practicalities won't concern us. It may be very difficult, for example, to count the number of grains of sand on a beach, or the number of molecules in our body, or even the number of quantum states of our brain; but, in principle, these things can be done. We're also not even concerned with whether the number of things to be counted is finite or not. Even if we have to go on counting \emph{forever}, if each element of a set can be given a number, then we are happy to call that set \emph{countable}.

You would think that most everything would be countable; but, unexpected things are provably not. For example, it's impossible to count the number of provable theorems in a mathematical theory. That's a theorem. Another thing that can't be counted is the number of degrees of freedom in a field --- like the fields that constitute all our most basic theories of physics. This fact causes headaches (which we will discuss soon) but, perhaps surprisingly, it doesn't prevent us from very accurately describing Nature. Indeed, it would appear that our basic understanding of physics and mathematics relies exclusively on things that cannot be counted. But are we right to think this, or is there something incomplete about our current understanding of physics? Are field theories really fundamental or do we need a new framework for making sense of our world? Another way to phrase this is ask whether the degrees of freedom that make up Nature are \emph{discrete}, like the bits in a computer, and can be counted with the natural numbers, or whether they are \emph{continuous} like the elements of the real line. We will present a scenario where physics must be continuous because, at a fundamental level, it is scale invariant. If scale doesn't exist, there can be no minimum length (because this minimum length would provide a preferred scale in the theory) and, therefore, no discreteness and no way to fundamentally capture the physics of our world on a standard computer. We will show how observers in Einstein's special theory of Relativity can be reinterpreted as observers in a scale-invariant space. This relationship appears to be intimately linked with a new formulation of gravity called \emph{Shape Dynamics} \cite{gryb:shape_dyn}, which we will come back to at the end. For the moment, we can look for a clue for how to make pragmatic progress on these issues by considering the nature of the gravitational force.

I already mentioned the infinity of degrees of freedom in a field theory and the headaches they cause for physicists and mathematicians. These headaches are most commonly dealt with using a framework called \emph{renormalization}. Renormalization works on the principle that a theory behaves in a different way depending on how accurate your measuring procedure is. Some theories, like the field theories that describe the forces important for atoms, nuclei, and nuclear constituents, behave in an increasingly simple way when the measuring procedure becomes more and more accurate. For these theories, a finite number of measurements need to be performed for the parameters of the theory (like the masses of the particles or the relative strength of the force) to be determined. These theories are called \emph{renormalizable} and are deemed acceptable field theories because, once the parameters have been obtained, the result of any measurement can be predicted once the accuracy of the measurement is specified. Unfortunately, the simplest quantum theory of gravity is not a theory of this type.

There is a simple way to understand why this is true. In General Relativity, energy warps spacetime and there is no limit to the amount of warping that is possible. This means that regions of very dense energy can collapse under their own weight to form regions of infinitely curved spacetime. Near these singular points, nothing can escape and these now familiar regions are called \emph{black holes}. In Quantum Theory, if you make a measurement of an object's position, its momentum inevitably becomes more uncertain. Thus, by accurately measuring the position of an object, you create a high probability that this object will have a correspondingly large energy. Now put General Relativity and Quantum Theory together. Doing this, you can reach a point where you can measure the position of a particle so accurately that its average energy becomes large enough to produce a black hole that is bigger than the region in which you are trying to detect that very particle. Gravity produces a black hole that ruins your quantum measurement. The length scale at which this happens is called the \emph{Plank scale} and understanding what happens to physics at these scales is one of the great mysteries of modern physics.

There are several strategies for attacking this problem. The two most common are listed below:
\begin{itemize}
    \item {\it Introduce new physics:} It may be that General Relativity is not the correct theory of gravity and that a new theory, with nicer quantum properties, takes over at length scales that we have not yet been able to probe experimentally.
    \item {\it Abandon the continuum:} If spacetime is fundamentally discrete --- that is, if there are only a countable number of degrees of freedom in the theory --- and this discreteness presents itself before the Plank scale is reached, then the theory is cured because the problematic region has been eliminated.
\end{itemize}
The two most studied modern approaches to quantum gravity, String Theory and Loop Quantum Gravity, make use of these strategies; the former using the first strategy and the latter the second. 

Although the research programs following these two strategies have made impressive progress, important open questions still remain. For example, approaches that try to introduce new physics inevitably run into the problem that General Relativity is a very robust theory, so that it is difficult to modify it without ruining its basic structures. One is then presented with many ambiguities for how to do this and these ambiguities are not easy to resolve without conflicting with known experiments. On the other hand, the ``fundamental discreteness'' scenario seems to suffer from a rather immediate drawback. Since no mathematical framework with a countable number of elements can ever be proven to be finished (because of the theorem I mentioned earlier), there is no way to prove that your ``fundamental'' theory is ever truly fundamental. It is impossible to know for certain whether some other theory is not underlying the true behaviour of the system. It is not known whether frameworks based on the continuum are subject to a similar restriction, but perhaps they can be proven to be superior in this regard.

Given these and other open questions in the standard approaches, it is perhaps justified to consider other strategies. One such strategy embraces the continuum and requires that physics should be fundamentally scale invariant at its most basic level. In approaches that follow this strategy, there can be no notion of discreteness because a minimum length scale would be quite obviously in conflict with the requirement that scale is meaningless.

\section{The Case for Scale Invariance}

In the theory of renormalization, when continuous fields are considered, one of the most common ways for a particular theory to be renormalizable is for it to be scale invariant at its most fundamental level --- that is, when infinitely accurate (or high energy) measurements are considered.\footnote{The technical requirement is that the theory have an \emph{ultra-violet fixed point}.} This makes sense because a scale invariant limit of this kind means that the theory itself eventually stops changing at a certain point as you keep making your measurements more precise. This provides you with a kind of anchor that allows you to determine what the theory should look like when you start making coarser and coarser measurements. If the anchor isn't fixed, the theory could drift anywhere. However, there is an even more basic reason for wanting fundamental scale invariance in your theory: only dimensionless quantities have objective meaning. A ``meter'' doesn't have any meaning on its own unless it is compared against the length of another object. Thus, it is only in the scale-invariant description of the theory that its parameters can be sensibly given an objective, dimensionless value.

For these reasons, there exists a number of approaches that aim to describe gravity in a scale-invariant way, either exactly or in some high energy limit of the theory. This task would seem difficult because scale seems to be an important part of our description of modern physics. Neither the Standard Model (our current framework for understanding the sub-atomic physics) nor General Relativity are manifestly scale invariant. Nevertheless, there are several approaches that aim to achieve a scale-invariant description of Nature. The most direct of these are approaches that aim to describe gravity and matter directly in terms of locally scale-invariant physics. Such approaches were originally pursued by Weyl \cite{weyl:conformal_1918} and recently by authors such as `t Hooft \cite{Hooft:2010nc} or Mannheim \cite{Mannheim:2011ds}. Other approaches aim to recover an approximate notion of scale invariance in the high energy limit of the theory\footnote{I am referring to the search for a UV fixed point (which has scale invariance in form of vanishing beta-functions) in the theory space of General Relativity}. These include the asymptotic safety program \cite{Percacci:2007sz}, which aims to make sense of quantum General Relativity as a quantum field theory. Unfortunately, despite many years of effort, none of these approaches have provided a completely adequate picture of scale-invariant gravity.

What all of these approaches have in common, is that they are considering a \emph{spacetime} notion of scale invariance. In this work, we will instead be concerned with a slightly different notion: that of \emph{spatial} scale invariance. We will now present a framework which suggests that \emph{spatial} scale invariance may actually be hidden in the framework of Special Relativity, which forms the starting point for General Relativity. Indeed, we will give an argument showing how the concepts of spacetime and spatial scale invariance can be interchanged using well-known mathematical transformations. This interchangeability seems to be intimately connected with a new formulation of General Relativity, called \emph{Shape Dynamics}, where local scale invariance is manifest.

\section{Scale Invariance in an Expanding Universe}

\subsection{The Expanding Universe}

That the spacetime description of Special Relativity can be traded for a scale-invariant description of separated space and time can be made possible by considering two ingredients:
\begin{itemize}
    \item[a.] That space is closed and has the shape of a 3 dimensional sphere. This means that an observer can head in the same direction and (eventually) come back to their original location.
    \item[b.] That space is expanding and that this expansion if fueled by a very small, positive parameter called the \emph{cosmological constant}, which we will discuss briefly.
\end{itemize}
The first ingredient is an assumption of simplicity. If true, then a holistic picture of the world is, at least in principle, possible. The second ingredient is taken from observation, and must be part of our description of reality. Note that what is actually observed as evidence of expansion is what is called the \emph{red shift}, which means that the frequency of light appears to get stretched out over time. We measure this by comparing \emph{ratios} of lengths, which gives a dimensionless (i.e., scale-invariant) number. Thus, the concept of expansion can be well-defined in a way that does make reference to any absolute scales.  It is this second, observationally motivated ingredient that will be key to our argument. This is a new input in that it was unknown to Einstein and others during the development of Relativity.

We will now add these two postulates to Einstein's original postulates of Special Relativity:
\begin{enumerate}
    \item The Laws of physics should take the same form for any \emph{inertial observer}.\footnote{An \emph{inertial observer} is one that is not moving under the influence of an external force.}
    \item The speed light is defined to be a finite constant, $c$, for all inertial observers.
\end{enumerate}
The first postulate is an assumption of simplicity, while the second is proposed for compatibility with observations. Just as in Special Relativity, we will assume the existence of idealized rods and clocks which can be perfectly synchronized. An inertial observer can label any event that can occur using a coordinate indicated by these idealized clocks and rods using some prescribed procedure (which won't interest us here). If space is not expanding, it is then an easy exercise to show, using Einstein's postulates, that different inertial observers will register a different set of coordinates for the same event and that these coordinates are related through a set of transformations called the \emph{Poincar\'e transformations}. We then say that the physical events are \emph{Poincar\'e invariant}.

Adding the two postulates [a] and [b] changes things considerably. We would like to imagine how the Universe would be expanding under the influence of the observed cosmological constant if the influences of all other forms of matter and energy could be ignored. This assumption is, in fact, not valid but we will still be able to apply our model to the real world if we take this as a \emph{local principle} for constructing a more realistic theory. That means that we should think of our ``expanding Universe'' below as what an observer would see in a sufficiently small region near the location of the observer.

We know what such an expanding Universe should look like because we understand how the Universe would expand in the presence of a cosmological constant only. What happens is as follows. Space has the shape of a 3 dimensional sphere, and at any time this sphere has some radius. To be precise, let's call that radius, $r$. In time, this radius changes its size. For an inertial observer who is stationary in some coordinate system, $r$ changes according to a very specific rule: the square of $r$ is equal to the square of the time interval read on the observer's clock (times $c$) plus the square of the cosmological horizon that we will call $\ell$ (and which is related to the cosmological \emph{constant} by taking its inverse square):
\begin{equation}\label{eq:ds hyp}
    r^2 = (ct)^2 + \ell^2.
\end{equation}

\begin{figure}
        \centering
        \begin{subfigure}[b]{0.52\textwidth}
                \centering
                \includegraphics[width=\textwidth]{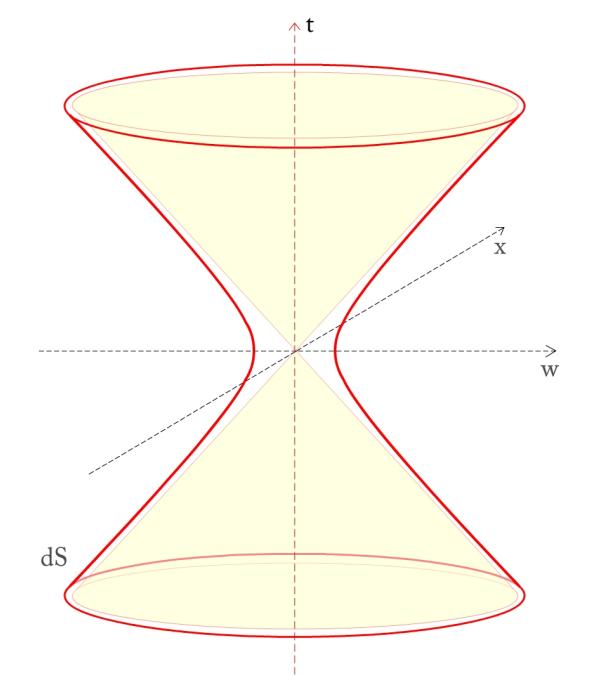}
                \caption{dS hyperboloid}
                \label{fig:dS hyperboloid}
        \end{subfigure}%
        ~ 
        \begin{subfigure}[b]{0.47\textwidth}
                \centering
		\includegraphics[width=\textwidth]{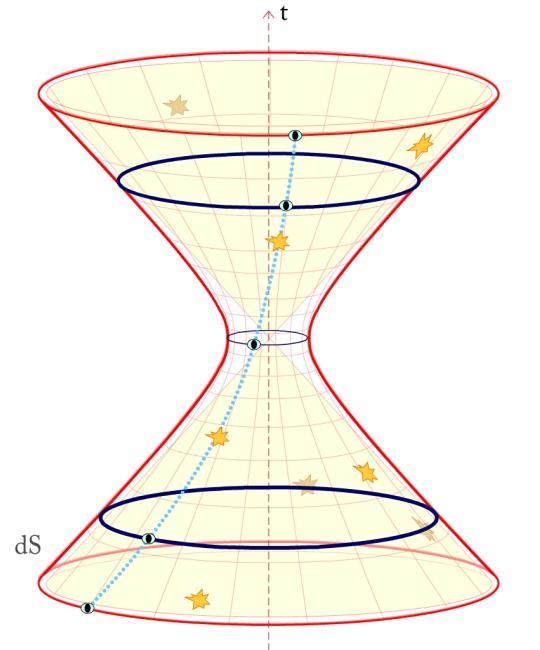}
		\caption{A typical inertial observer.}\label{fig:inertial observer}
        \end{subfigure}
        ~ 
          \caption{The de~Sitter hyperbola. Observers are depicted as eyes and events as bursts of light.}
\end{figure}

The resulting spacetime is curved and has the shape of a hyperboloid. We will call this a \emph{de~Sitter (dS)} hyperboloid after the first person to study its properties. We can visualize a curved dS hyperboloid by drawing it inside of a flat spacetime in one extra dimension. This is completely analogous to how one can visualize a 2 dimensional curved sphere by drawing it inside a 3 dimensional flat space, even though the third dimension is not accessible to observers confined to live on the surface of the sphere. In Figure~\ref{fig:dS hyperboloid}, we show what the dS hyperboloid looks like. Time flows upwards. Since, for visualization purposes, we have added an extra spatial dimension, these are labeled by $(x,y,z,w)$ and we have collapsed the $(x,y,z)$-direction and have only shown the $x$-direction. 
These spatial coordinates should not be confused with the three \emph{real} spatial coordinates $(\theta, \phi, \psi)$ which are periodic and can be chosen to represent angles on a 3 dimensional sphere. Just as observers on a sphere cannot move into the regions interior and exterior of the sphere, an observer in dS space cannot move into the regions interior and exterior to the hyperboloid. As can be seen from Figure~\ref{fig:dS hyperboloid}, in the infinite past, space has a larger and larger size. This shrinks down to the minimum value of $\ell$ before beginning an expansion phase that continues into the infinite future.
An important feature of the dS hyperboloid are its inertial observers. They follow the ``straight lines'' on the hyperboloid, which are represented by hyperbola that extend from the distant past to the distant future. A typical observer of this kind is illustrated in light blue in Figure~\ref{fig:inertial observer}.


A \emph{stationary} observer, is an observer whose $(\theta, \phi, \psi)$-coordinates on the sphere do not change in time, $t$. We know, however, from the fact that space is expanding, that the fictitious $(x,y,z,w)$-coordinates must change in time in order for the relation \eqref{eq:ds hyp} to be maintained. By convention, we can pick the $w$-direction to be the only direction that is changing. Thus, for the stationary observer, $x=y=z=0$. Events which occur in the dS hyperboloid are distinguished by points to which the stationary observer will attribute a particular set of coordinates. Other inertial observers are ``straight lines'' on the hyperboloid in the sense that they follow the extreme paths on the hyperboloid.

Now we are in a position to understand how Einstein's theory of Special Relativity can be extended to an expanding Universe. We want to be able to relate the coordinates attributed to events in the stationary observer's reference frame to the coordinates attributed to the \emph{same} events in the reference frame of some other inertial observer. We are thus looking for the set of transformations that generalize the Poincar\'e transformations in a dS Universe. In order for Einstein's first principle to hold, these transformations must preserve the shape of the dS hyperboloid so that the form of the Laws of physics are unchanged. In other words, we are looking for the set of symmetry transformations of the dS hyperboloid. Mathematically, this corresponds to the set of transformations that preserve the relation \eqref{eq:ds hyp} because this is the defining relation of the hyperboloid. Since the hyperboloid itself is unchanged, the distance between two points remains the same so that inertial observers will continue to be inertial observers after the transformations.

It is now a rather straightforward mathematical exercise to identify the set of transformations that preserve the form of \eqref{eq:ds hyp}. Let's refer to them as the group of dS symmetries. In total, there are 10 of them (the same number as the original Poincar\'e transformations in flat space $(x,y,z,t)$-spacetime). There are 6 attributed to the symmetries of the 3 dimensional sphere: 3 of which are rotations and 3 are translations of the $(\theta, \phi, \psi)$ coordinates. These are analogous to the translations and rotations in the familiar flat $(x,y,z)$-space that we learn about in high-school. The other 4 are associated with time. One is just a time translation. The remaining three are what happens when you change your velocity in either of the three possible directions $(\theta, \phi, \psi)$. These are called \emph{boosts}. In the next section, we will try to break these down in more detail and then show how they can be related to a different kind of symmetry: scale invariance in space.

\subsection{Breaking Down the Symmetries}

In order to describe the symmetries of the expanding Universe, we will try to draw pictures that will help us understand the beautiful structure of the mathematics. In order to achieve conceptual clarity and in order for us to actually be able to draw things on paper, we will often have to suppress certain dimensions. It will then require a bit of an imagination to get the full picture of what is going on.

\begin{figure}
        \centering
        \begin{subfigure}[b]{0.4\textwidth}
                \centering
                \includegraphics[width=\textwidth]{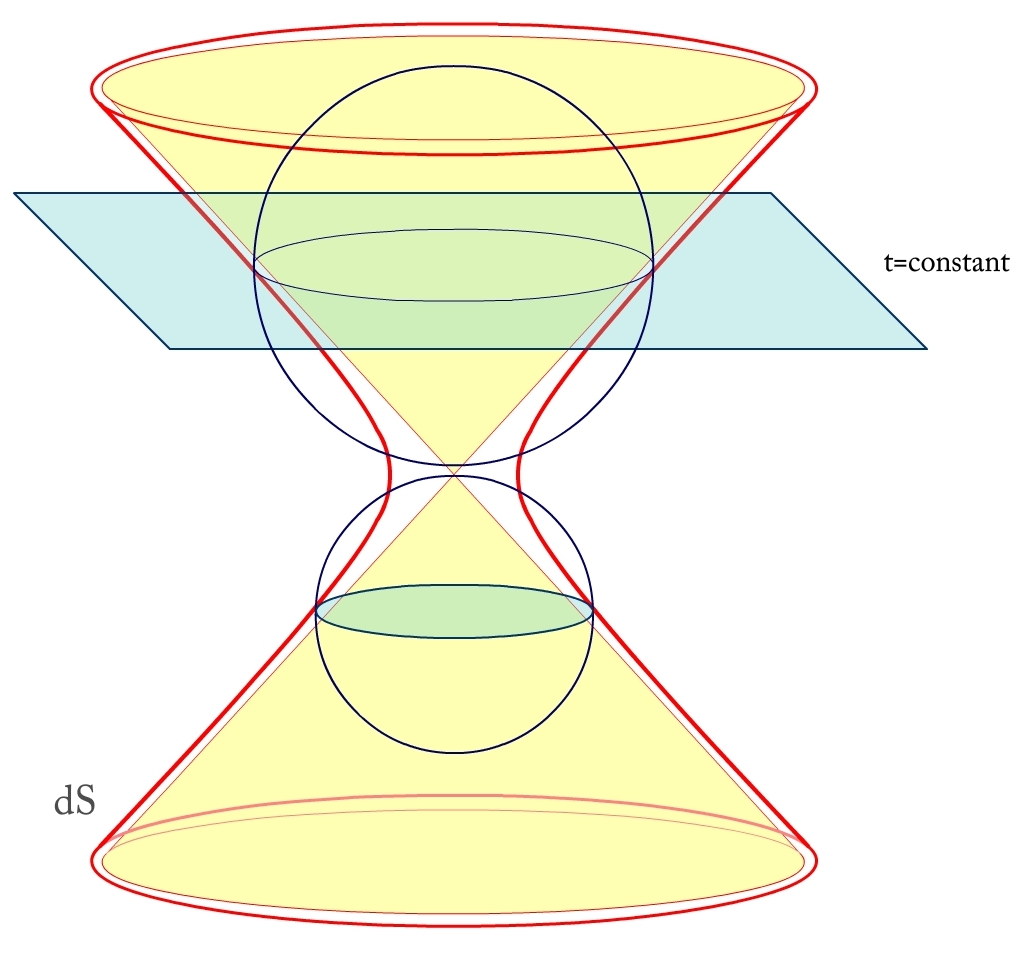}
                \caption{Planes of constant $t$.}
                \label{fig:constant t}
        \end{subfigure}%
        ~ 
        \begin{subfigure}[b]{0.59\textwidth}
                \centering
                \includegraphics[width=\textwidth]{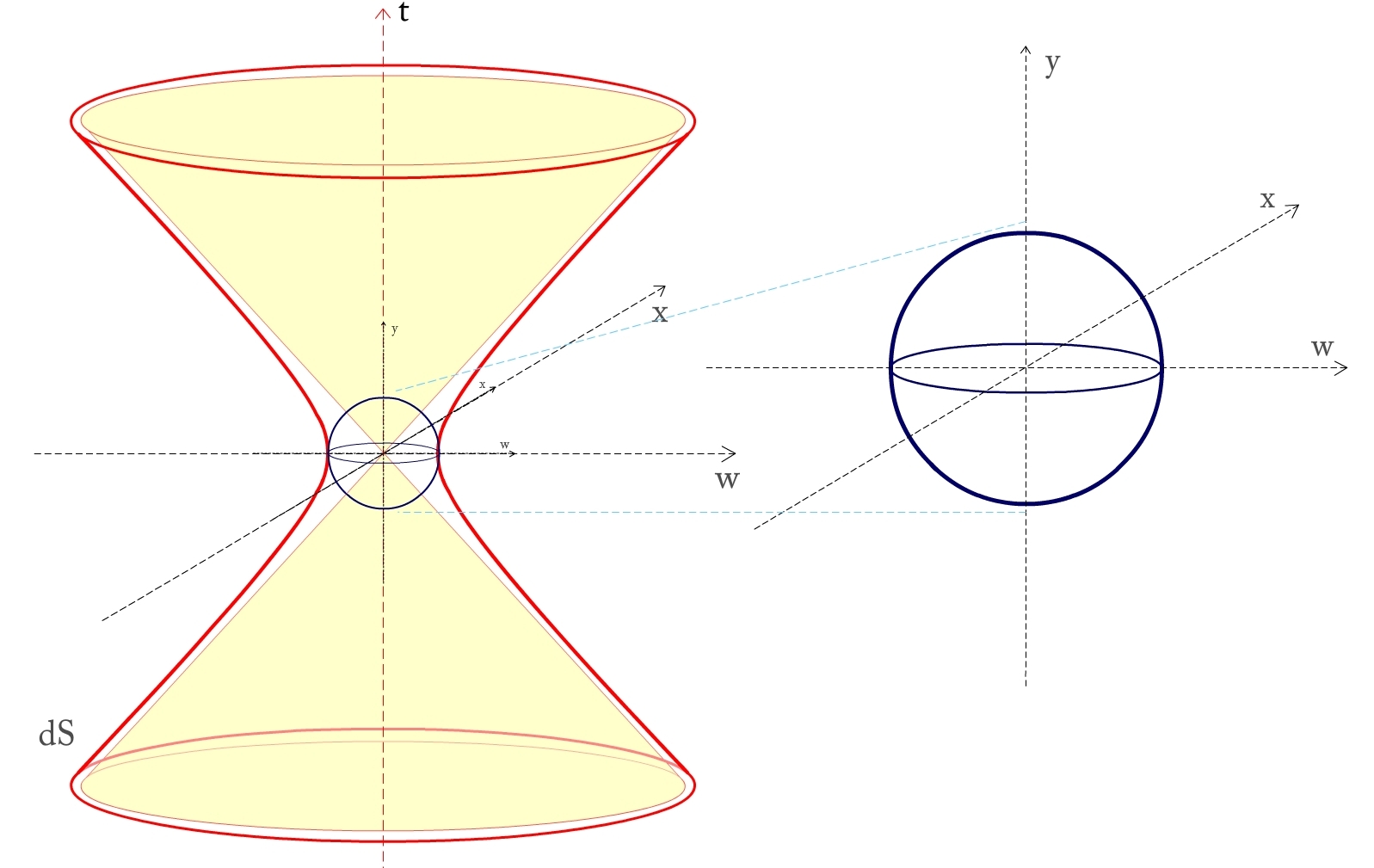}
                \caption{2d spheres.}
                \label{fig:2 spheres}
        \end{subfigure}
        ~ 
          \caption{The planes of constant $t$ intersect the hyperboloid in 2d spheres.}
\end{figure}

To make our task easier we will, from now on, suppress the $z$-direction completely. There are now 3 spaces of interest (and later, we will add a fourth): the fictitious extra dimensional flat space, which now has the 4 coordinates $(x,y,w,t)$; the dS hyperboloid, which is our model for the Universe and has the 3 coordinates $(\theta, \phi, t)$; and the 2 dimensional sphere with coordinates $(\theta,\phi)$. Here are how these spaces are related. As already discussed, the dS hyperboloid is the surface in the fictitious flat space whose points obey the relation \eqref{eq:ds hyp}. The 2-spheres can be obtained by drawing surfaces of constant $t$. These are the planes drawn in Figure~\ref{fig:constant t} and they intersect the dS hyperboloid in what look like circles.
However, these are not actually circles because we've suppressed one spatial dimension. Remember that the $r$-direction is actually the $xy$-plane. This means that, what looks like circles, are actually 2 dimensional spheres, which, unfortunately, we can't draw in our limited number of dimensions. Figure~\ref{fig:2 spheres} shows how we can imagine the 2-spheres which intersect the planes of constant $t$. These are ``snapshots'' of the Universe and they are clearly changing in size over time.

\begin{wrapfigure}{L}{0.55\textwidth}
  \vspace{-20pt}
  \begin{center}
    \includegraphics[width=0.54\textwidth]{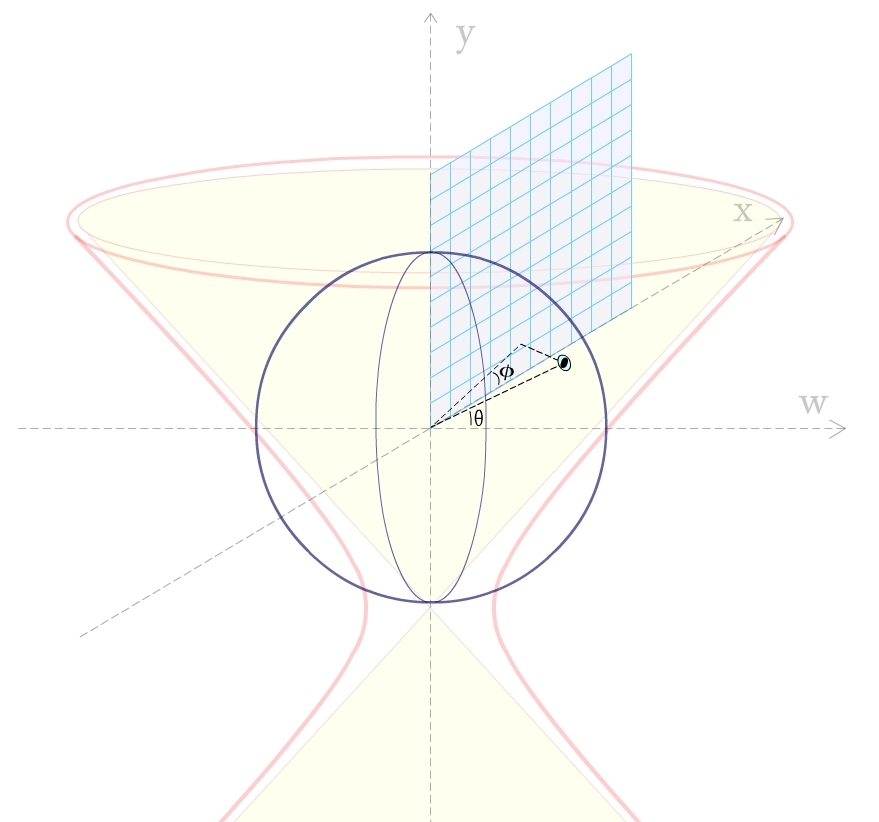}
  \end{center}
    \caption{The polar angles $\theta$ and $\phi$ on the 2-sphere.}\label{fig:polar coords}
\end{wrapfigure}

Lets examine these 2d spheres. Remember our convention where we called the $w$-direction the direction where our stationary observer is moving. This suggests that we choose a spherical coordinate system where the $w$-direction points up. It is then customary to choose two angles, $\theta$ and $\phi$, that parameterize the 2d sphere. The $\phi$ variable goes from $0$ to $2\pi$ and represents the angle in the $xy$-plane. The $\theta$ variable goes from $0$ to $\pi$ and represents the angle to the $w$-axis (as is shown in Figure~\ref{fig:polar coords}). We can see that we now have consistency with our postulate~[a] since, as advertised, our $(\theta,\phi)$ variables are periodic
\begin{align}
    \theta &= \theta + 2\pi & \phi &= \phi + 2\pi.
\end{align}

We can now visualize the symmetries of the dS hyperboloid. The simplest ones are the ones associated with the symmetries of the 2-sphere. Since we are suppressing one dimension, we now have 3: one rotation and two translations. The rotation is the familiar rotation of the sphere that keeps the North pole fixed; i.e., it is a rotation about the $w$-axis. The two translations are the two different ways to move the North pole. These involve independent or simultaneous shifts of the $\theta$ and $\phi$ coordinates.

The symmetries associated with time are a bit harder to visualize, especially because the size of the spheres is changing in time. It is perhaps simplest to visualize by how they act on the hyperboloid in the flat, higher dimensional space. In this space, they look like a kind of time-space rotation of the hyperboloid around one of the spatial axes. Take, for example, the $tw$-``rotation'' of the hyperboloid about the $xy$-axis. This is easiest to visualize from the point of view of the stationary observer at $x=y=0$. This observer follows the hyperbola given by the intersection of the $xy$-plane with the dS hyperboloid. Thus, the $tw$-``rotation'' just pushes the observer up in time along its trajectory. The only effect of this transformation is for the 2-spheres to change their size, like what happens in Figure~\ref{fig:constant t}. This will be an important fact later. To get the remaining two symmetries, we just have to replace the $w$-axis with the $x$ or $y$ axes. Now, however, these transformations do not look like simple time translations for the stationary observer, but involve increasing the velocity in the $\theta$ and $\phi$ directions (this obviously changes the definition of our stationary observers).

Those are the symmetries of the dS hyperboloid. They are the generalizations of the usual Poincar\'e transformations of flat spacetime to the case of a Universe expanding due to a cosmological constant. Our next, and final, task is to show that these transformations can be recast as a set of scale invariant transformations that, in particular, preserve the shape of the configurations of observers in the Universe (in a precise way that we will specify below). For this we will need to introduce one last space on which our ``shapes'' will eventually live. This is just a flat Euclidean plane. In general, it can have $(X,Y,Z)$ coordinates (were we used capitals to distinguish them from the coordinates in the fictitious flat spacetime), but, for simplicity, we will suppress the $Z$-dimension.

\begin{wrapfigure}{R}{0.5\textwidth}
  \vspace{-20pt}
  \begin{center}
    \includegraphics[width=0.45\textwidth]{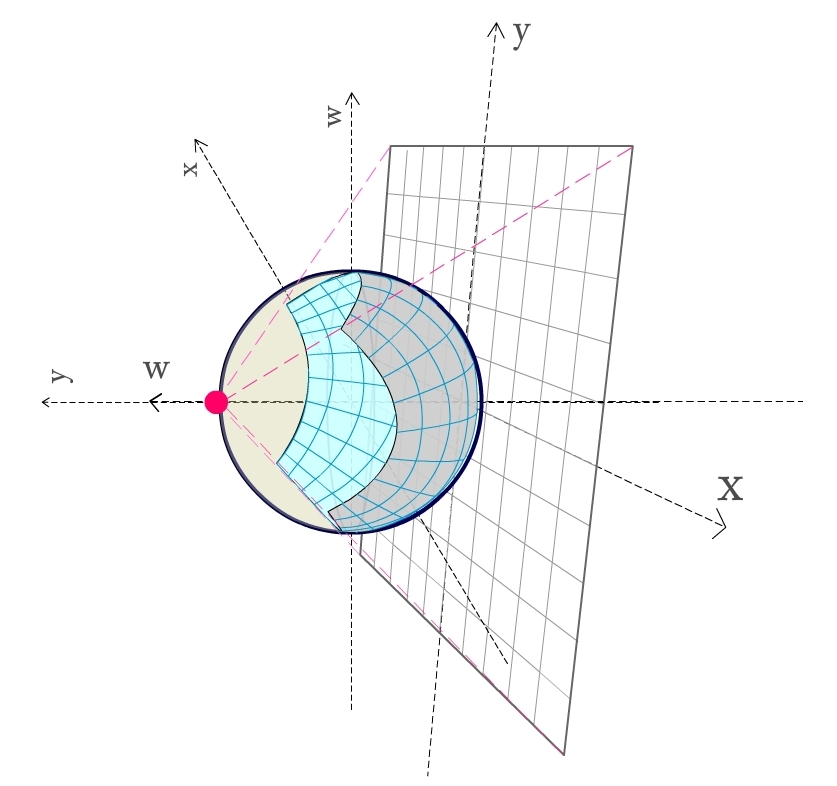}
  \end{center}
    \caption{The stereographic projection of a grid onto the plane.}\label{fig:stereoproject}
\end{wrapfigure}

What we will now require is a way to map the points on the 2-sphere to the points of this flat 2-plane. There are many ways of doing this but the one we will be interested in has been used by map makers for ages. This is because map makers have the same problem as us: they have to project locations on the round Earth to points on a flat map. The technique used by navigators is called the \emph{stereographic projection} and the same property that makes it useful for navigation will also be useful for us: it preservers angles! The way to perform the stereographic projection is to imagine a light bulb siting on the South pole of the sphere (i.e., the point $x=y=0$ and $w= -\sqrt{\ell^2 + (ct)^2}$). Then, imagine placing the plane so that it is tangent to the sphere at the North pole (see Figure~\ref{fig:stereoproject}).
The shadow cast on the plane by a point on the sphere is its stereographic projection. It's clear that any point on the sphere (except the South pole) will have a stereographic projection onto the plane.

We now return to the key property of the stereographic projection: any angle formed by the intersection of two lines on the sphere will be preserved by the projection. Consider a particular $t= \text{const}$ 2-sphere in a model Universe where a stationary observer is making observations in the presence of two other inertial observers. At this instant, one can draw imaginary lines between each of the particles forming a kind of triangle. Under the stereographic projection, the angles of this triangle are preserved. Because an observer can only makes measurements locally, the angles they measure are the only objective way for them to determine the ``shape'' of this three particle configuration. We can then say that the shape is preserved under the projection.

By ``scale invariance'', we mean that the theory doesn't depend on the size of the configurations of the system. Instead, only the shapes, as defined above, should be important. This kind of scale invariance is also called \emph{conformal invariance}, and the transformations that preserve angles (or the shapes of the instantaneous configurations of the system) are called \emph{conformal transformations}. Again, it is a relatively straightforward mathematical exercise to determine what these conformal transformations are. In 2 dimensions, they can be written mathematically using the complex variable
\begin{equation}
    \zeta = X + i Y,
\end{equation}
where $i = \sqrt{-1}$. In terms of this variable, the conformal transformations are also called \emph{M\"obius transformations}. The mathematical definition of the M\"obius transformations is given in Appendix~\ref{apx:Mobius}. Here we will describe them physically.

\begin{wrapfigure}{R}{0.5\textwidth}
  \vspace{-20pt}
  \begin{center}
    \includegraphics[width=0.45\textwidth]{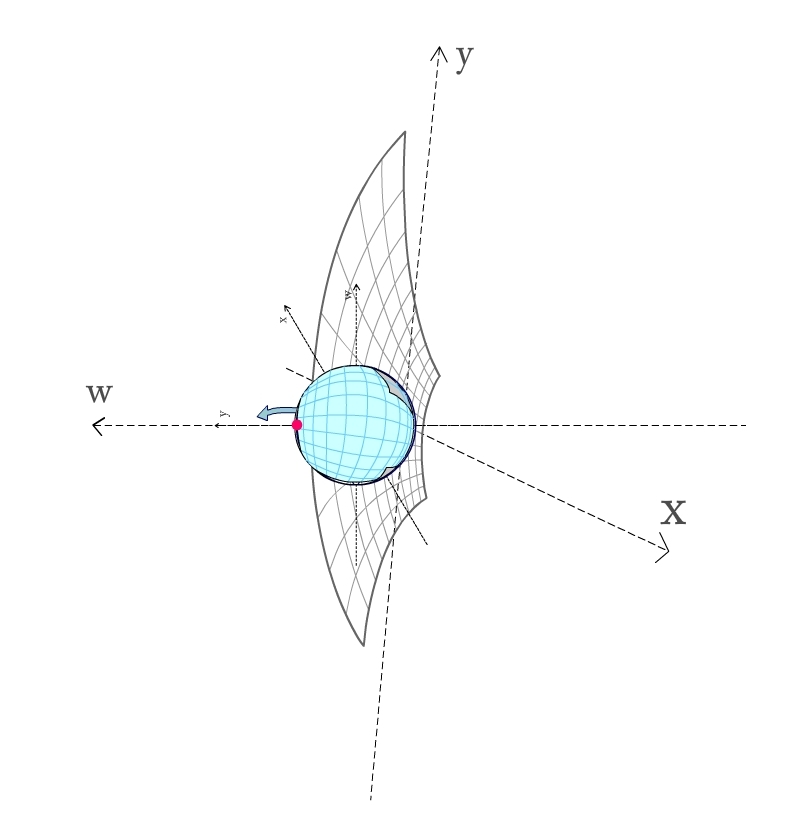}
  \end{center}
  \vspace{-20pt}
    \caption{An inversion.}\label{fig:inversion}
    \vspace{-10pt}
\end{wrapfigure}

There are six different kinds M\"obius transformations. The simplest involve no changes of scale at all. There are three of these. The first, is a rotation in the $XY$-plane and the other two are translations in the $X$ and $Y$ directions. Clearly, these won't change the shape of the system. The other three involve changes in the global scale. The first of these are dilatations where only the global scale of the system is changed. The last two are a bit harder to visualize. They are called \emph{special conformal transformations} and can be visualized most easily by imagining how they can be stereographically projected onto the plane from the sphere. A special conformal transformation is a combination of translating the position of the sphere over the plane and performing an \emph{inversion}, which involves rotating the position of the North pole as shown in Figure~\ref{fig:inversion}. In fact, all the M\"obius transformations can be represented in a simple way using the stereographic projection.

One may have noticed a similarity between the symmetry group of the dS Universe and the M\"obius transformations. In both cases, there are six transformations and, in both cases, they can be represented by how they act on surfaces of constant $t$. Indeed --- and this is the key observation necessary for our analysis --- the M\"obius transformations can be shown to be equivalent to the dS symmetries when the expansion of space starts to become large! This happens during the very early and very late times of our Universe. Thus, the behaviour of inertial observers during these epochs can be used to link the spacetime description of events to a scale-invariant one. Using this correspondence, we can map a system of inertial observers in the dS Universe to conformally invariant point particles (with a fixed time parametrization) in flat space. A concrete prescription for doing this is given in Appendix~\ref{apx:model}. Many more details about the model presented here can be found in the technical paper \cite{Gryb:2014iva} which feeds heavily off of the mathematical structures discussed in \cite{Wise:2013uma}.

\begin{figure}
        \centering
        \begin{subfigure}[b]{0.47\textwidth}
                \centering
                \includegraphics[width=\textwidth]{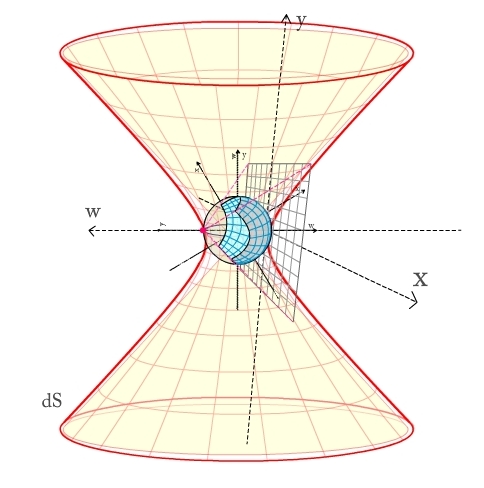}
                \caption{The project at time $t$.}
                \label{fig:dil1}
        \end{subfigure}%
        ~ 
        \begin{subfigure}[b]{0.52\textwidth}
                \centering
                \includegraphics[width=\textwidth]{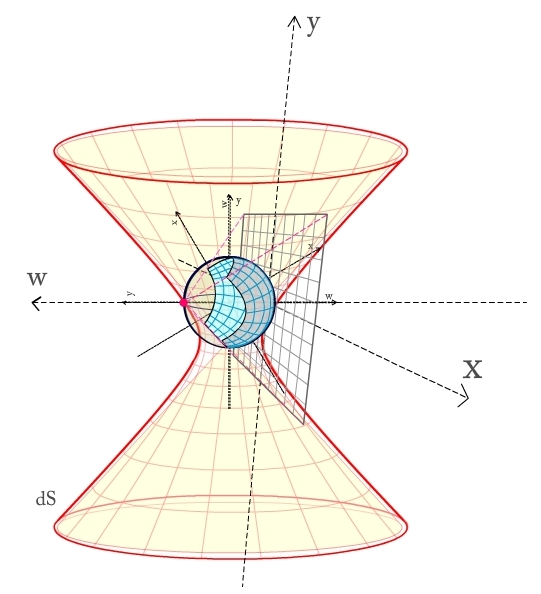}
                \caption{The project at a later time.}
                \label{fig:dil2}
        \end{subfigure}
        ~ 
          \caption{In time, the projection of the grid grows in size.}\label{fig:dilatations}
\end{figure}

This entire construction relies upon being able to represent the dS symmetries, in the distant past and future, as conformal transformations on the plane via the stereographic projection. We will now illustrate how this can be done for the two simplest of these transformations. For a brief description of this, consult Appendix~\ref{apx:model} or, for the full computation, see \cite{Gryb:2014iva}. The simplest transformations to relate are the rotations around the $w$-axis in dS space. These are quite obviously equivalent to rotations in the $XY$-plane after stereographic projection. The second transformation, which is slightly more non-trivial, is the time translation for the stationary observer. As we pointed out, this transformation corresponds to simple time translations for the stationary observer. Since the size of the 2-spheres changes in time, these correspond to dilatations on the $XY$-plane. Figure~\ref{fig:dilatations} shows how this happens. The remaining M\"obius transformations are harder to visualize because they are combinations of translations and boosts. Furthermore, these can only be shown to be equivalent in the distant future and distant past (this leads to the \emph{holographic} nature of the construction presented in the Appendix~\ref{apx:model}). We encourage the reader to try to work these out for themselves.

\section{Conclusion}

We have shown how it is possible to reinterpret the trajectories of inertial observers in dS space in terms of the trajectories of particles where only the instantaneous shape of the spatial configurations count. The (asymptotic) symmetries of dS space can be transformed to the conformal symmetries on the plane via a stereographic projection. The picture we describe here could help clarify how to understand the meaning of inertial observers in Shape Dynamics, which is a new framework for gravity where \emph{local} scale is traded for the time part of spacetime symmetries.

In this essay, it has been the \emph{global} aspects of scale invariance that have concerned us. However, in a full theory of gravity, it is \emph{local} Poincar\'e, and the corresponding \emph{local} scale, transformations that are relevant. For that more complicated case, one can build up a curved spacetime by gluing together local patches of flat spacetime using the techniques of \emph{Cartan geometry} \cite{Sharpe:Cartan_geometry}. Could it be possible to do something similar in Shape Dynamics; i.e., glue together local patches of \emph{conformally} flat spaces that could then be related to General Relativity through the correspondence outlined here? A first step towards doing this was performed in $2+1$ dimensions \cite{gryb:2_plus_1}, but many additional difficulties arise in the more physical case of $3+1$ dimensions. Some further investigations towards this end have been explored by Derek Wise in \cite{Wise:2013uma}. It is clear that further insights are needed to sort out these intriguing possibilities, but the relationships described here could be a first step towards achieving such insights.

But what does all of this suggest? Perhaps it suggests that there is a way to think of quantum gravity in fully scale-invariant terms. If true, this would provide a new mechanism for being able to deal with the uncountably infinite number of degrees of freedom in the gravitational field without introducing discreteness at the Plank scale. It would give us a new view of the continuum, where the infinitely large can exist in the infinitely small, possible because scale is a matter of your point of view, not a matter of fact. Like observers on Escher's Circle Limit III, we can continue to peer into the infinite complexities of our world, ever pondering the mysteries that lie beyond.
\begin{figure}
        \centering
	    \includegraphics[width=.65\textwidth]{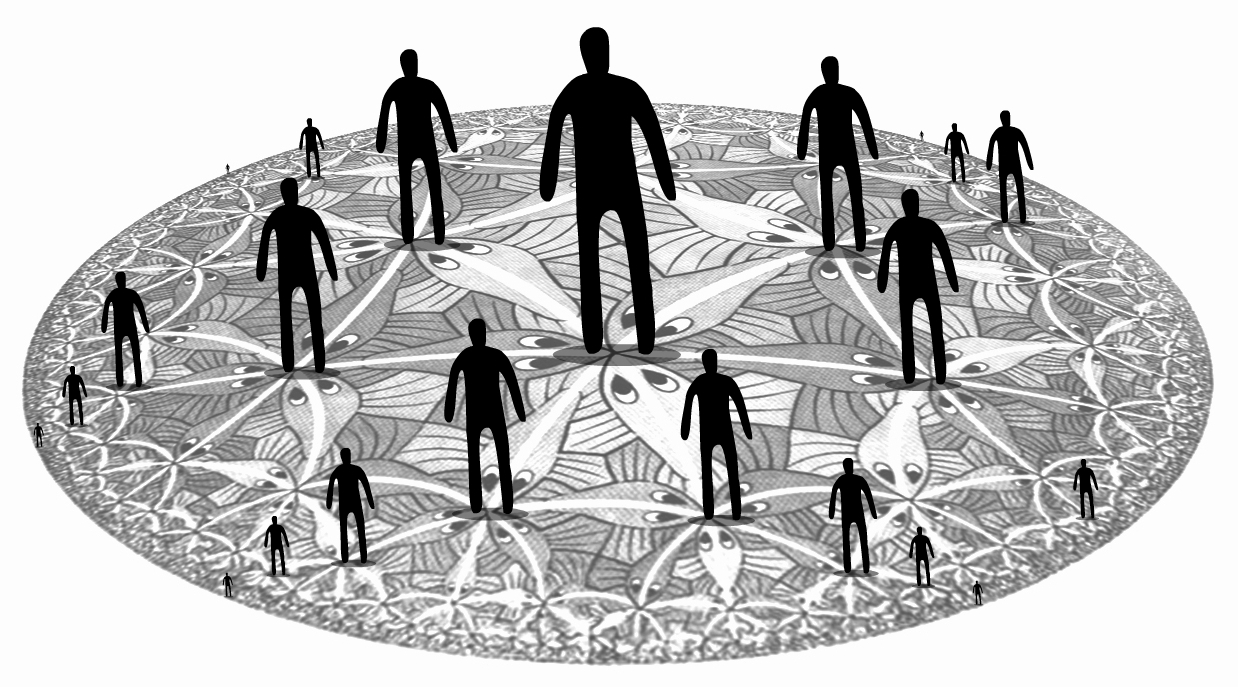}
	    \caption{Escher's Circle Limit III.}
	    \label{fig:escher}
\end{figure}

\clearpage
\appendix

\section{M\"obius Transformations and the Lorentz Transformations}
\label{apx:Mobius}

The M\"obius transformations are defined as:
\begin{equation}
    \zeta \to \frac{a\zeta + b}{c\zeta + d},
\end{equation}
where $a,b,c,d$ are complex numbers obeying $ac - bd \neq 0$. This group is well-known to be isomorphic to the projective special linear group PSL$(2,\mathbbm C)$, which, in turn, is isomorphic to the orthochronous Lorentz group SO$^+(3,1)$. It is this property that we exploit in Appendix~\ref{apx:model}. For more info on the M\"obius transformations and for visualizations which inspired our diagrams on stereographic projection, see \cite{Arnold:mobius}.

\section{de~Sitter Inertial Observers to Scale Invariant Particles}
\label{apx:model}

For a much more detailed account of the material presented here, see the technical paper \cite{Gryb:2014iva}.

We are inspired by the Shape Dynamics formulation of gravity, as presented in \cite{gryb:shape_dyn}, where equivalence with GR is manifest in Constant Mean Curvature (CMC) slicings of solutions to the Einstein equations. For dS$^{d,1}$ spacetime, the CMC slices are constant $t$ hypersurfaces in the ambient $\mathbbm R^{d+1,1}$ and have $\mathcal S^{d}$ topology. To see this, we can use a convenient choice of coordinates for the embedding:
\begin{align}
    t &= \ell \sinh \varphi & x^I &= \ell \cosh \varphi\, \tilde x^I,
\end{align}
where $I = 1\hdots (d+1)$ and $\tilde x^I \tilde x^J \delta_{IJ} = \tilde x^2 = 1$. Using these coordinates, the induced metric is
\begin{equation}
    ds^2 = -\ell^2 d\varphi^2 + \ell^2 \cosh^2\varphi\, d\Omega^2,
\end{equation}
where $d\Omega^2$ is the line element on the unit $d$-sphere. Since the spatial metric is conformal to the metric on the unit sphere (which is homogeneous), it is clear that this slicing must be CMC.

We now consider a useful set of coordinates:
\begin{align}
    x^\pm &= x^0 \pm x^{d+1} & X^i &= \frac{x^i}{x^0 - x^{d+1}}\,,
\end{align}
where $i = 1 \hdots d$. The $x^\pm$ are just light-cone coordinates in the ambient space. We can single out one of these, namely $x^-$, as a convenient time variable and write the other $x^+ = \frac{x^2 - \ell^2}{x^0 - x^{d+1}} = \frac 1 {x^-} \lf( \frac{X^2}{(x^-)^2} - \ell^2\rt)$ using the definition of de~Sitter spacetime. The $X^i$'s are a convenient choice of spatial coordinates because, as can be shown with a straightforward calculation, in the limit as $t \to \pm\infty$ (i.e., the conformal boundary of spacetime), they are just giving the stereographic projection of coordinates on the constant-$t$ hypersurfaces onto a Euclidean plane:
\begin{equation}
    X^i \to \frac{\tilde x^i}{1 - \tilde x^{d+1}}\,.
\end{equation}
The utility of these coordinates becomes obvious when one considers the action of the ambient Lorentz transformations $x^\mu \to \Lambda^\mu_\nu x^\nu$ on the new coordinates. Indeed, near the conformal boundary, it can be shown that $x^- \to x^-$ and that the $X^i$ transform under the full conformal group.

This last property allows us to define a scale-invariant theory holographically using the action principle for massive particles following bulk geodesics. To see how this can be done, consider the action for a single particle of mass $m$ following a geodesic in dS 
\begin{equation}
    S(X^i_{in}, X^i_{out}) = \lim_{t_0 \to \infty} \int_{-t_0}^{t_0} d t \lf[ m \sqrt{ - \eta_{\mu\nu} \dot x^\mu \dot x^\nu} + \lambda \lf( \eta_{\mu\nu} x^\mu x^\nu - \ell^2 \rt)  \rt]\,,
\end{equation}
where $X^i_{in}$ and $X^i_{out}$ are the asymptotic values of the coordinates $X^i$ on the past and future conformal boundary. The Lagrange multiplier $\lambda$ enforces the constraint keeping the particle on the dS hyperboloid. If we evaluate this along the classical solution while carefully taking the limit, $S$ becomes of a function of the asymptotic values of $X^i$. Moreover, as was just indicated, it is also conformally invariant. This means that it can be interpreted as the Hamilton--Jacobi function of some holographically defined conformally invariant theory.

In \cite{Gryb:2014iva}, $S$ is explicitly computed in this limit. The result is
\begin{equation}
    S = \frac {m\ell}2 \lf[ \ln\lf( \frac{(X_{in} - X_{out})^2}{\epsilon^2} - 2 \rt)  + \mathcal O(\epsilon^4) \rt]\,,
\end{equation}
where $\epsilon = \ell/t \to 0$ as $t \to \infty$. This behaves exactly like the Hamilton--Jacobi functional of a reparametrization invariant theory with potential equal to $V = \frac 1 {X^2}$, which is well-known to be scale invariant. We see that a free massive particle in dS spacetime can be equivalently described by a scale-invariant particle in a reparametrization invariant theory. Furthermore, the bulk dS isometries map explicitly to conformal transformations in the dual theory, as advertised.

\clearpage

\bibliographystyle{utphys}
\bibliography{mach}

\end{document}